\documentclass{ws-ijmpb}

\usepackage{graphics}      
\usepackage{graphicx} 
\usepackage{url}           
\usepackage{hyperref}
 \hypersetup{colorlinks=true, urlcolor=blue, 
linkcolor=blue, citecolor=blue} 
\usepackage{lineno}
\usepackage{indentfirst}
\usepackage{bm}
\usepackage{dsfont}
\usepackage{mathrsfs}
\usepackage{amsmath}
\usepackage{amssymb}
\usepackage{leftidx}

\begin{document}

\markboth{H. ARIAN ZAD AND H. MOVAHHEDIAN}
{Classical correlation and Quantum Entanglement in the Mixed-Spin Ising-XY Model}

%
\catchline{}{}{}{}{}
%

\title{CLASSICAL CORRELATION AND QUANTUM ENTANGLEMENT IN THE MIXED-SPIN ISING-XY MODEL WITH DZYALOSHINSKII-MORIYA INTERACTION}

\author{HAMID ARIAN ZAD}

\address{Young Researchers and Elite Club, Mashhad Branch, Islamic Azad University, Mashhad, Iran
$^{*}$arianzad.hamid@mshdiau.ac.ir} 

\author{HOSSEIN MOVAHHEDIAN}

\address{Department of Physics, Shahrood University of Technology,
3619995161, Shahrood, Iran}

\maketitle


\begin{abstract}
In the present work, initially a mixed-three-spin (1/2,1,1/2) cell of a mixed-$N$-spin chain with Ising-XY model is introduced, for which pair spins (1,1/2) have Ising-type interaction and pair spins (1/2,1/2) have both XY-type and  Dzyaloshinskii-Moriya(DM) interactions together. An external homogeneous magnetic field $B$ is considered for the system in thermal equilibrium. Integer-spins have a single-ion anisotropy property with coefficient $\zeta$. Then, we investigate the quantum entanglement between half-spins (1/2,1/2), by means of the concurrence. Classical correlation(CC) for this pair of spins is investigated as well as the concurrence and some interesting the temperature, the magnetic field and the DM interaction properties are expressed. Moreover, single-ion anisotropy effects on the correlation between half-spins is verified. According to the verifications based on the communication channels category by D. Rossini, V. Giovannetti and R. Fazio \cite{Rossini}, we theoretically consider such tripartite spin model as an ideal quantum channel, then calculate its information transmission rate and express some differences in behaviour between this suggested model and introduced simple models in the previous works(chains without spin integer and DM interaction) from information transferring protocol point of view.
\end{abstract}


\section{Preliminaries}
Heretofore, there are a lot of interests to investigate the various correlations (whether quantum or classical) for an ideal system \cite{Ribeiro,A.Abliz1,Sarandy1,A.Abliz2,Wells}. If we would like to verify the quantum correlation between parts of a system then may be bound us to investigate the entanglement. Quantum entanglement is a special property which can exist only in the quantum systems \cite{Horodecki2005,Horodecki2007,Horodecki2009,Cerf1}. Thereby, most of researchers confine themselves to verify the entanglement to understand the behaviour of such systems in the various situations \cite{Cerf1,Cerf2,Cerf3,Horodecki1998,McCulloch,Langari1,Saif}. In this way, spin models are ideal candidates for generating and manipulating of entangled states and for studying the entanglement \cite{Sarandy1,Wootters,Vedral,Kamta,Zhang,Xiaoguang1,Rigolin1} by verification some stimulating quantities such as, the concurrence \cite{Wootters,Vedral,Kamta}, negativity \cite{Vidal,XWang,Arian2}, quantum discord \cite{Zurek,Modi,Sarandy2,Sarandy3,Xiaoguang2,Arian1}, quantum disorder \cite{Grande}, correlation functions \cite{Sarandy1,Sarandy3} and von Neumann entropies \cite{Cerf3,Neumann,Cerf4,Wilde}. Somewhere, spin models have been studied with the DM interaction \cite{A.Abliz2,Dzyaloshinskii,Moriya,Cao2,Cao3,Cao4}, that such interaction arises naturally in the perturbation theory due to the spin-orbit coupling in magnetic systems. 

Straightforward researches have been caried out to investigate interaction between the next-nearest-neighbour sites of a Heisenberg spin model in Refs. \cite{Chen,Haibin,Shyiko,Kwek}. Such interaction may has an essential role to generate a Heisenberg model with diamond chain topology by organizing  mixture of particles that have different spins.
Motivated by this issue, several studies have been done on the mixture of different spins with various models and many interesting results have been reported \cite{Ivan,Yamamoto,Langari2}. Diamond chains as attractive structures among these spin models were exactly investigated from quantum entanglement, quantum correlation, phase transitions etc.  view points \cite{Ivanov2,Rojas1,Ananikian,Seyit,Canova,Rojas2}. 

 The motivation for the study of a diamond chain with the Ising-XXZ model  is that it can describes real materials such as  natural mineral azurite $Cu_3(CO_3)_2(OH)_2$, where according to experimental results, theoretical calculations are interestingly reasonable in this case \cite{Kikuchi}(another polymeric coordination compounds such as $M_3(OH)_2$ with spin-1 Heisenberg diamond chain were investigated in the literature \cite{Tong}). Another quantum spin models consisting of diamond-shaped cells can be theoretically suggested and solved. In this regard, we here are interested to introduce a few body diamond chain with specific model and verify its bipartite CC and also entanglement in the some physical situations.
 
 In our previous works \cite{Arian2,Arian1}, we analyzed bipartite quantum entanglement in the mixed-three-spin system (1/2,1,1/2) with two different \textquoteleft XXX Heisenberg\textquoteright and \textquoteleft Ising-XY\textquoteright  models  in the vicinity of an external homogeneous magnetic field. This paper has been devoted to verify CC and the quantum entanglement between half-spins (1/2,1/2) of same as the second model for which an additional DM interaction is considered between half-spins. Some interesting temperature,the magnetic field, the DM interaction and the other applied coefficients properties especially single-ion anisotropy related to the integer spin are expressed. The main purpose of this work is to provide the exact solution for the generalized version of the mixed spin-1/2 and spin-1 Ising-XY diamond chain, which should bring a deep insight into how the thermal and the magnetic properties depend on the spins-1/2 and spin-1 of the model, in order words, we are going to understand the spin-1 existence has how much physical effects on the correlation between spins-1/2. 

 Forthermore, the suggested model is considered as a memoryless communication channel between hypothetical sender Alice and the receiver Bob, then quantum information transmission rate $\mathcal{R}$ \cite{Fazio1,Demianowicz,Burgarth1} is numerically verified. The ratio $\mathcal{R}$ describes the maximum number of qubits one can transfer through the channel per unit of time. Before, we proved that similar mixed-spin model can be considered as a communication channel for transferring qutrits \cite{Zad3}.

The paper is organized as follows. In the next section, we first characterize the concurrence as a measure of entanglement and CC between the spins (1/2,1/2), also we have a quick look at the model as a communication channel(Sec. \ref{channel}). In Sec. \ref{mixed} we define our favorite model with an analytical Hamiltonian and get its eigenvectors and eigenvalues. Then, we extract density matrix of the bipartite spins (1/2,1/2) from density matrix of the mixed-three-spin system in representation of the basis states. In Sec. \ref{numerical}, we show the numerical calculations and simulations of the concurrence and CC between the spins (1/2,1/2), with respect to the temperature, the magnetic field, the coupling constant $J$, the single-ion anisotropy $\zeta$, the DM interaction $D$  and the anisotropy parameter $\gamma$ associated to the XY interaction. Also, information transmission rate of the mixed-three-spin chain channel is theoretically investigated. Section \ref{conclusions} is devoted to discussions and a summary of conclusions.
\section{Introduction to the Concurrence and the Classical Correlation(CC)}\label{introduction}
\subsection{Concurrence}
The concurrence that is a measure of entanglement, can be defined for bipartite spin systems as
\begin{equation}
\mathcal{C}_{12}(\rho)=max\{0,2\lambda-\sum^{4}_{i=1}\lambda_i\},
\end{equation}
where $\lambda=\max\{\lambda_1,\lambda_2,\lambda_3,\lambda_4\}$ and $ \lambda_i$  are square roots of the eigenvalues of the inner product
\begin{equation}
R=\rho\tilde{\rho},
\end{equation}
with
\begin{equation}
\tilde{\rho}=(\sigma_y\otimes\sigma_y){\rho^\dagger}(\sigma_y\otimes\sigma_y),
\end{equation}
where in the basis states $\{\mid{00}\rangle,\mid{01}\rangle,\mid{10}\rangle,\mid{11}\rangle\}$, the density matrix of a quantum system with Hamiltonian $H$ in thermal equilibrium is defined as
\begin{equation}
\rho_{eq}=\frac{\exp (-\beta H)}{Tr[\exp(-\beta H)]},
\end{equation}
where $\beta=1/T$(we set $k_B=1$) in which $T$ is the temperature and $Z=Tr[\exp(-\beta H)]$ is the partition function of the system. ${\rho^\dagger}$ denotes the complex conjugation of the density matrix $\rho$ \cite{Wootters,Vedral} and
\begin{equation}
\sigma_y = \left(
\begin{array}{cc}
0 & -i \\
i & 0 \\
\end{array} \right).
\end{equation}

Hitherto, the concurrence was explicitly calculated and simulated in terms of the temperature, the magnetic field, the DM interaction  etc. for the various spin models. In the some of references cited here and references therein, it has been mentioned that $\mathcal{C}(\rho)$ behaves as sudden death at striking critical points, which is called \textquotedblleft entanglement sudden death\textquotedblright(also review Refs. \cite{Ann,Yu}). In the following, we investigate the concurrence changes with respect to the temperature, the magnetic field, the anisotropy coefficient $\gamma$, the single-ion anisotropy $\zeta$ and the DM interaction $D$. Moreover, we would like know, that what is the effect of spin-1 existence in the system on the temperature, the magnetic field and the DM interaction dependences of the concurrence corresponding to the spins (1/2,1/2), consequently we extract some interesting outcomes.

\subsection{Classical correlation}\label{CC}
We here recall the concept of CC for the spins (1/2,1/2) briefly. Total correlation in a bipartite system formed by (sub)systems $\mathscr{A}$ and $\mathscr{B}$ in a composite Hilbert space ${H}_{bi}={H}^{\mathscr{A}}\otimes {H}^{\mathscr{B}}$ is quantified by the quantum mutual information \cite{Cerf1,Cerf2,Horodecki1998} as 
\begin{equation}
\mathcal{I}(\rho_{\mathscr{A}}:\rho_\mathscr{B})=S(\rho_\mathscr{A})+S(\rho_\mathscr{B})-S(\rho_{\mathscr{A}\mathscr{B}}),
\end{equation}
where, $S(\rho)=-Tr[\rho \log_2(\rho)]$  is the von Neumann entropy in which $\rho_{\mathscr{A(B)}}=Tr_{\mathscr{A(B)}}(\rho)$. The quantum mutual information includes quantum information and classical one (see  Refs. \cite{Cerf1,Barnett}).  After a measurement on one of the (sub)systems such as $\mathscr{A}$, the amount of information obtained about the another (sub)system $\mathscr{B}$ is defined as CC. CC can be defined in terms of POVM measurement \cite{Sarandy1,Wells,Sarandy3}. Let us consider a set of projective measurements $\lbrace \mathscr{B}^\kappa \rbrace $ performed locally only on part $\mathscr{B}$ then, the probability of measurement outcome $\kappa$ is defined as
\begin{equation}
p_{\kappa}=Tr_{\mathscr{A}\mathscr{B}}[(I^\mathscr{A}\otimes \mathscr{B}^\kappa)\rho_{\mathscr{A}\mathscr{B}}(I^\mathscr{A}\otimes \mathscr{B}^\kappa)],
\end{equation}
where $I^\mathscr{A}$ denotes the identity operator for the (sub)system $\mathscr{A}$. After this measurement, state of the subsystem $\mathscr{A}$ is described by the conditional density operator
\begin{equation}
\rho_{\kappa}=\frac{1}{p_\kappa}[(I^\mathscr{A}\otimes \mathscr{B}^\kappa)\rho_{\mathscr{A}\mathscr{B}}(I^\mathscr{A}\otimes \mathscr{B}^\kappa)].
\end{equation}
The projectors $\mathscr{B}^\kappa $ can be characterized as $\mathscr{B}^{\kappa} =V \Pi^{\kappa} V^\dagger$ where, $\Pi^{\kappa}=\vert \kappa \rangle \langle \kappa \vert$ at which $\kappa=\{0,1\}$. We parametrize the matrix $V$ as
\begin{equation}
V = \left(
\begin{array}{cc}
cos(\frac{\theta}{2}) & e^{-i\phi} sin(\frac{\theta}{2})\\
 e^{i\phi} sin(\frac{\theta}{2}) & -cos(\frac{\theta}{2}) \\
\end{array} \right),
\end{equation}
where $V\in U(2)$, $0\leq\theta\leq\pi$ and $0\leq\phi\leq2\pi$. We define the suprimum of the difference between the von Neumann
entropy $S(\rho_\mathscr{A} )$ and the based-on-measurement(POVM) quantum conditional entropy $ S(\rho_{\mathscr{A}\mathscr{B}}\vert\lbrace \mathscr{B}^{\kappa}\rbrace)=\sum_{\kappa}p_{\kappa}S(\rho_{\kappa})$ of the subsystem $\mathscr{A}$ as
\begin{equation}\label{CCEq}
CC(\rho_{\mathscr{A}\mathscr{B}})=sup_{\{ \mathscr{B}^{\kappa}\}}\{S(\rho_\mathscr{A})-S(\rho_{\mathscr{A}\mathscr{B}}\vert \lbrace \mathscr{B}^{\kappa}\rbrace)\},
\end{equation}
where $S_{min}(\rho_{\mathscr{A}\mathscr{B}})=min_{\{\mathscr{B}^{\kappa}\}}S(\rho_{\mathscr{A}\mathscr{B}}\vert\lbrace \mathscr{B}^{\kappa}\rbrace)$. CC has been precisely verified in Ref. \cite{Modi}.

\section{Channel Capacity}\label{channel}
 Study of the classical information and communication channels was first characterized by Shannon \cite{Shannon} and its quantum analogous was promoted by von Neumann \cite{Neumann} and more studies have been devoted to these regimes in the last decades \cite{Cerf4,Demianowicz,Rossini,Holevo1,Bennett,Cerf5,Lloyd1,Benenti}. The information sent via quantum communication channels is carried by quantum states(qubits), classical information(bits) can also be transmitted through quantum channels, namely, any channel that is able to transmit quantum information can be likewise used for transmitting classical information. One can find profound detections about quantum communication channels category in Refs. \cite{Benenti,Caruso}.

Recently, it was proposed for using simple spin chains with specific models as communication channels \cite{Fazio1,Fazio2,Burgarth,Arshed1}. Some interesting suggested models as communication channel are included Pauli channels \cite{Palma1,Arshed2}, depolarizing channels, dephasing channels \cite{Arshed1}, spin chains channels \cite{Fazio1,Fazio2}, electromagnetic channels \cite{Lloyd2} and Gaussian channels \cite{Holevo1,Palma2,Lee}. Channel capacity is the maximum rate of a communication channel which information can be reliably carried. Hence, researchers are interested to study on the capacity of a channel with memory or memoryless for transmitting or storing unknown quantum states \cite{Caruso}.

\begin{figure}
\begin{center}
\includegraphics[width=11.3cm,height=3.3cm]{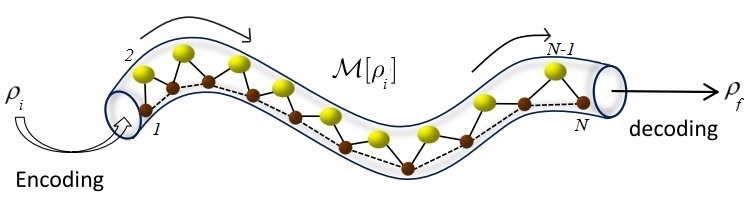}
\caption{Caricature of the transferring information through a mixed-spin chain as a quantum communication channel.}
\label{fig:map}
\end{center}
\end{figure}

 When an arbitrary state $\rho$ is propagated through a communication channel with capacity, it can be wholly characterized by designing mapping protocol as the bellow form
\begin{equation}
\mathcal{M}: \rho_{i}\longrightarrow \rho_{f}=\mathcal{M}[\rho_{i}],
\end{equation}
where $\rho_{i}$ is the initial state and $\rho_{f}$ is the mapped state through channel. Such mapping is performed by unitary  transformation operator related to the feature of the channel as
\begin{equation}
\mathcal{M}[\rho]=U\rho U^\dagger.
\end{equation}
Caricature of this protocol is shown in Fig. \ref{fig:map}.

This paper puts substantial limits on the amount of information that can be transmitted reliably along a mixed-three-spin chain memoryless channel. The action of a generic quantum channel  denoted on a single system as $\mathcal{E}_1$ can be defined as $\mathcal{E}_N=\mathcal{E}_1^{\otimes N}$, where $N$ represents the number of channel uses. The quantum capacity $\mathcal{Q}$ measured in qubits per channel use, is defined as 
\begin{equation}\label{QC}
\mathcal{Q}=\max\limits_{N\rightarrow \infty}\frac{\mathcal{Q}_N}{N},
\end{equation}
where $\mathcal{Q}_N=\max_{\rho}\big[\mathcal{S}(\mathcal{E}_N(\rho))-\mathcal{S}_e(\rho,\mathcal{E}_N)\big]$, which denotes the maximum coherent information. $\mathcal{S}$ represents the von Neumann entropy, and $\mathcal{S}_e$ is the entropy exchange, namely $\mathcal{S}_e(\rho,\mathcal{E})=\mathcal{S}\big((\mathds{1}_{\mathcal{H}}\otimes \mathcal{E})(\mid\Psi_{\rho}\rangle\langle\Psi_{\rho}\mid)\big)$, and state $\mid\Psi_{\rho}\rangle$ is any purification of $\rho$ by means of a reference quantum system $\mathcal{H}$, i.e. $\rho=Tr_{\mathcal{H}}\big[\mid\Psi_{\rho}\rangle\langle\Psi_{\rho}\mid)\big]$.

\begin{figure}
\begin{center}
\includegraphics[width=10cm,height=4.5cm]{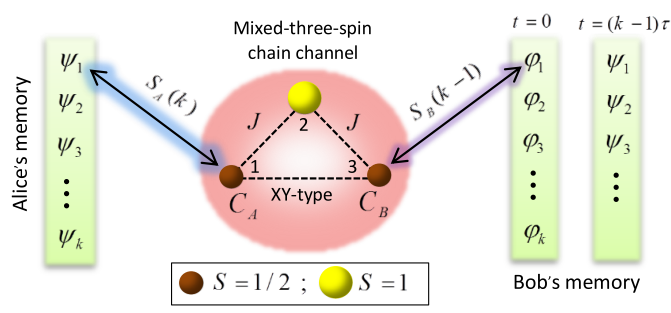}
\caption{Schematic representation of the mixed-three-spin chain as a quantum communication channel.}
\label{fig:spinchain}
\end{center}
\end{figure}

The transfer protocol can be as the follows: (I) sender (Alice) applies a SWAP operation $S_A$ for the set of unknown states $\vert\psi_n\rangle_A=\mid{\psi_n,\cdots\psi_3,\psi_2,\psi_1}\rangle$, and first part of the spin chain $C_A$ (see Fig. \ref{fig:spinchain}). Here, it is assumed that the spin chain is initially in a separable state $\mid{\psi}\rangle_C= \mid{\downarrow ,\circlearrowleft,\downarrow}\rangle$ which, here states of spins-half and spin-integer of the tripartite system are set up in the $S_z$ and $ J_z $ down states $\mid{0}\rangle=\mid{-1/2}\rangle=\mid{\downarrow}\rangle$ and $\mid{-1}\rangle=\mid{\circlearrowleft}\rangle$ respectively, (II) after time $t=\tau$, the receiver (Bob) tries to recover the information sent from Alice by means of some decoding protocols, namely, Bob will extract information sent by applying a SWAP operator $S_B$. This operator couples Bob$^,$s memory with $C_B$ which corresponds to the reduced density operator for the $N$-th qubit. In fact, after time $t=\tau$, this protocol can be described as a completely positive trace preserving (CPTP) mapping from input state density matrices to output state density matrices, in the form 
\begin{equation}
\rho_A\longrightarrow \mathcal{M}[\rho_A]=Tr^{T_B}[U(\rho_A\otimes \rho_C)U^\dagger],
\end{equation}
where $\mathcal{M}$ represents the mapping scenario of a real memoryless channel and $U$ is the unitary transformation which describes joint evolution of the composite Hamiltonian $H_{AC}=H^A\otimes H^C$, $Tr^{T_B}$ denotes to trace over all spins except receiver B. 

 In the present work, it is considered a scenario where the sender and the receiver use their spins belong to the spin chain for encoding and decoding information(classical or quantum). From communication theory point of view, this consideration is not enough persuasive, but on the one hand, the consequences can be treated analytically.
 
For doing transmission protocol, suppose that Alice has a memory with state $\vert \Psi\rangle_A=\cdots \vert \psi_3\rangle\otimes\vert \psi_2\rangle\otimes\vert \psi_1\rangle$, where $\vert \psi_i\rangle=\alpha^{\prime}_i\vert \downarrow\rangle +\beta^{\prime}_i\vert \uparrow\rangle$($i=\{1,2,3,\cdots\}$ and $|\alpha^{\prime}_i|^2+|\beta^{\prime}_i|^2=1$). By starting the transmission protocol($t=0$) by Alice via coupling first memory element $\vert \psi_1\rangle$ with state of the first chain spin $C_A$ through SWAP gate $S_A(1)$, memory element $\vert \psi_1\rangle$ will replace with the state of $C_A$. We here assumed that the mixed-three-spin chain be initially in state $\vert \downarrow \circlearrowleft \downarrow\rangle$.  
This procedure can be characterized as
\begin{equation}\label{memory1}
\begin{array}{lcl}
\big(\cdots \vert \psi_3\rangle\otimes\vert \psi_2\rangle\otimes\vert \psi_1\rangle\big)_A\otimes \vert \downarrow\circlearrowleft\downarrow\rangle_C\otimes \big(\vert \downarrow\rangle\otimes\vert \downarrow\rangle\otimes \vert \downarrow\rangle\cdots\big)_B\xrightarrow{S_A(1)}\\
\big(\cdots \vert \psi_3\psi_2 \downarrow\rangle\big)_A\big(\alpha^{\prime}_1\vert \downarrow\circlearrowleft\downarrow\rangle +\beta^{\prime}_1\vert \uparrow\circlearrowleft\downarrow\rangle\big)_C\big(\vert \downarrow\rangle\otimes\vert \downarrow\rangle\otimes \vert \downarrow\rangle\cdots\big)_B.
\end{array}
\end{equation}
After time evolution $\tau$, the first memory element $\vert \psi_1\rangle$ embedded in $C_A$ after first SWAP $S_A(1)$, spreads along the  mixed-three-spin chain, thereby, the total state (\ref{memory1}) becomes
\begin{equation}\label{memory2}
\begin{array}{lcl}
\big(\cdots \vert \psi_3\psi_2 \downarrow\rangle\big)_A\otimes\big(\big(\alpha^{\prime}_1\vert \downarrow\circlearrowleft\downarrow\rangle +\beta^{\prime}_1\Upsilon_{13}(\tau)\vert \downarrow\circlearrowleft\uparrow\rangle\big)_C\otimes\big(\vert \downarrow \downarrow \downarrow\cdots\rangle\big)_B,
\end{array}
\end{equation}
where
\begin{equation}\label{Upsilon}
\begin{array}{lcl}
\Upsilon_{13}(\tau)=\leftidx _{C}\langle\downarrow\circlearrowleft\downarrow\vert e^{\frac{-iH\tau}{\hbar}}\vert \downarrow\circlearrowleft\uparrow\rangle _C
\end{array}
\end{equation}
which is the probability amplitude of finding the spin up ($\vert\uparrow\rangle$) in the 3-th part of the mixed-three-spin chain($C_B$ presented in Fig. \ref{fig:spinchain}). In Sec. \ref{numerical}, we will use these statements to verify the transmission rate of the channel numerically.

\section{Suggested Spin Model and Theoretical Background}\label{mixed}
We introduce Hamiltonian of the mixed-$N$-spin system with Ising-XY model (mixed-spin chain shown in Fig. \ref{fig:map}) which is in an external homogeneous magnetic field $B$, as the follows
\begin{equation}\label{hamiltonian}
\begin{array}{lcl}
H=\sum\limits_{i=1}^M\big((1+\gamma) {S}^{x}_{i,1}{S}^{x}_{i,3}+(1-\gamma) {S}^{y}_{i,1}{S}^{y}_{i,3}\big) \\
+\sum\limits_{i=1}^{M}\vec{B}\cdot(\vec{S}_{i,1}+\vec{S}_{i,3})+\sum\limits_{i=1}^{M}\vec{D}\cdot(\vec{S}_{i,1}\times\vec{S}_{i,3}) \\
+\sum\limits_{i=1}^{M}J({S}^{z}_{i,1}{J}^{z}_{i,2}+{J}^{z}_{i,2}{S}^{z}_{i,3})
+\sum\limits_{i=1}^{M}\zeta({J}^{z}_{i,2})^2
+\sum\limits_{i=1}^{M}\vec{B}\cdot\vec{J}_{i,2},
\end{array}
\end{equation}
where $i$ denotes the number of mixed-three-spin cell in the chain, $\gamma$ is anisotropy parameter, $J$ is the Ising coupling between the spins (1,1/2), $D$ is the DM interaction between spins-half of the cell and $\zeta$ is the single-ion anisotropy parameter considered for spins-integer. $\vec{S}=\lbrace {S}^x, {S}^y, {S}^z \rbrace$ and $\vec{J}=\lbrace {J}^x, {J}^y, {J}^z \rbrace$ are spin operators(with $\hbar=1$) which, are introduced as the following matrices
  \begin{equation} 
\begin{array}{lcl}
 S^x = \frac{1}{2} \left(
\begin{array}{cc}
0 & 1 \\
1 & 0 \\
\end{array} \right),
{S^y} = \frac{1}{2} \left(
\begin{array}{cc}
0 & -i \\
i & 0 \\
\end{array} \right),
{S^z} =\frac{1}{2} \left(
\begin{array}{cc}
1 & 0 \\
0& -1 \\
\end{array} \right),
\end{array}
\end{equation}
\begin{equation}
\begin{array}{lcl}
{J^x} = \frac{1}{\sqrt{2}} \left(
\begin{array}{ccc}
0 & 1 & 0\\
1 & 0 & 1\\
0 & 1 & 0 \\
\end{array} \right), 
{J^y} = \frac{1}{\sqrt{2}} \left(
\begin{array}{ccc}
0 & -i & 0\\
i & 0 & -i\\
0 & i & 0 \\
\end{array} \right),
{J^z} =\left(
\begin{array}{ccc}
1 & 0 & 0\\
0 & 0 & 0\\
0 & 0 & -1 \\
\end{array} \right).
\end{array}
\end{equation}

We here consider $M=1$ and $\vec{B}=B_z$ and $\vec{D}=D_z$, which are homogeneous magnetic field and DM interaction in the $z$-direction. Note that here all of introduced parameters are considered dimensionless parameters.
Eigenvectors of the Hamiltonian of the mixed-three-spin (1/2,1,1/2) chain are given by
\begin{equation}\label{eigenvectors}
\begin{array}{lcl}
\mid{\phi_1}\rangle= a\mid{\uparrow,\bigcirc,\downarrow}\rangle + \mid{\downarrow,\bigcirc,\uparrow}\rangle,\quad
\mid{\phi_2}\rangle= b\mid{\uparrow,\bigcirc,\downarrow}\rangle + \mid{\downarrow,\bigcirc,\uparrow}\rangle, \\
\mid{\phi_3}\rangle= e\mid{\uparrow,\circlearrowright,\uparrow}\rangle +\mid{\downarrow,\circlearrowright,\downarrow}\rangle,\quad
\mid{\phi_4}\rangle= f\mid{\uparrow,\circlearrowright,\uparrow}\rangle +\mid{\downarrow,\circlearrowright,\downarrow}\rangle,\\
\mid{\phi_5}\rangle= g\mid{\uparrow,\bigcirc,\uparrow}\rangle + \mid{\downarrow,\bigcirc,\downarrow}\rangle,\quad
\mid{\phi_{6}}\rangle= h\mid{\uparrow,\bigcirc,\uparrow}\rangle + \mid{\downarrow,\bigcirc,\downarrow}\rangle,\\
\mid{\phi_7}\rangle= j\mid{\uparrow,\circlearrowleft,\downarrow}\rangle +\mid{\downarrow,\circlearrowleft,\uparrow}\rangle,\quad
\mid{\phi_8}\rangle= k\mid{\uparrow,\circlearrowleft,\downarrow}\rangle +\mid{\downarrow,\circlearrowleft,\uparrow}\rangle,\\
\mid{\phi_9}\rangle= l\mid{\uparrow,\circlearrowleft,\uparrow}\rangle + \mid{\downarrow,\circlearrowleft,\downarrow}\rangle,\quad
\mid{\phi_{10}}\rangle= m\mid{\uparrow,\circlearrowleft,\uparrow}\rangle + \mid{\downarrow,\circlearrowleft,\downarrow}\rangle,\\
\mid{\phi_{11}}\rangle= n\mid{\uparrow,\circlearrowright,\downarrow}\rangle + \mid{\downarrow,\circlearrowright,\uparrow}\rangle,\quad
\mid{\phi_{12}}\rangle=  o\mid{\uparrow,\circlearrowright,\downarrow}\rangle + \mid{\downarrow,\circlearrowright,\uparrow}\rangle,
\end{array}
\end{equation}
where
\begin{equation}\label{coefficients}
\begin{array}{lcl}
a=-\frac{i(i-D)}{\sqrt{1+D^2}},\quad b=\frac{i(i-D)}{\sqrt{1+D^2}},\\
e=-\frac{1}{2}\frac{\gamma}{B+\frac{{\zeta}}{2}-\frac{1}{2}\sqrt{({\zeta}+2B)^2+4J({\zeta}+J+2B)+\gamma^2}+J}, \\
f=-\frac{1}{2}\frac{\gamma}{B+\frac{{\zeta}}{2}+\frac{1}{2}\sqrt{({\zeta}+2B)^2+4J({\zeta}+J+2B)+\gamma^2}+J}, \\
g=-\frac{1}{2}\frac{\gamma}{-\frac{1}{2}\sqrt{\gamma^2+4B^2}+B}, 
h=-\frac{1}{2}\frac{\gamma}{\frac{1}{2}\sqrt{\gamma^2+4B^2}+B}, \\
j= -\frac{i(i-D)}{-\zeta +\sqrt{1+\zeta^{2}+D^2}},\quad k=-\frac{i(i-D)}{-\zeta -\sqrt{1+\zeta^{2}+D^2}},\\
l=\frac{1}{2}\frac{\gamma}{\frac{\zeta}{2}-B+\frac{1}{2}\sqrt{(\zeta-2B)^2+4J(\zeta+J-2B)+\gamma^2}+J}, \\
m=\frac{1}{2}\frac{\gamma}{\frac{\zeta}{2}-B-\frac{1}{2}\sqrt{(\zeta-2B)^2+4J(\zeta+J-2B)+\gamma^2}+J},\\
n=\frac{i(i-D)}{-\zeta -\sqrt{1+\zeta^{2}+D^2}},\quad o=\frac{i(i-D)}{-\zeta +\sqrt{1+\zeta^{2}+D^2}},\\
\end{array}
\end{equation}
and the corresponding eigenvalues are
\begin{equation}\label{eigenvalues}
\begin{array}{lcl}
E_1={\zeta}+\frac{1}{2}\sqrt{1+D^2},\quad E_2={\zeta}-\frac{1}{2}\sqrt{1+D^2} \\
E_{3}=B+\frac{\zeta}{2}+\frac{1}{2}\sqrt{(\zeta+2B)^2+4J(\zeta+J+2B)+\gamma^2}, \\
E_{4}=B+\frac{\zeta}{2}-\frac{1}{2}\sqrt{(\zeta+2B)^2+4J(\zeta+J+2B)+\gamma^2}, \\
E_5={\zeta}+\frac{1}{2}\sqrt{\gamma^2+4B^2},\quad E_6={\zeta}-\frac{1}{2}\sqrt{\gamma^2+4B^2} \\
E_7=\frac{\zeta}{2}-B+\frac{1}{2}\sqrt{\zeta^2+D^2+1},\quad E_8=\frac{\zeta}{2}-B-\frac{1}{2}\sqrt{\zeta^2+D^2+1} \\
E_9=\frac{\zeta}{2}-B+\frac{1}{2}\sqrt{(\zeta-2B)^2+4J(\zeta+J-2B)+\gamma^2}, \\
E_{10}=\frac{\zeta}{2}-B-\frac{1}{2}\sqrt{(\zeta-2B)^2+4J(\zeta+J-2B)+\gamma^2}, \\
E_{11}=\frac{\zeta}{2}+B+\frac{1}{2}\sqrt{\zeta^2+D^2+1},\quad E_{12}=\frac{\zeta}{2}+B-\frac{1}{2}\sqrt{\zeta^2+D^2+1} \\
\end{array}
\end{equation}
In the basis states representation, the total density matrix of the considered tripartite system which is a thermal equilibrium state can be characterized by using latest equations. Hence, the density matrix of the pair spins (1/2,1/2) can be expressed as
\begin{equation}\label{density matrices}
\mathbf{\rho^{T_2}_{13}} =\frac{1}{Z}\left(
\begin{array}{cccc}
 \delta & 0 & 0 & \varsigma\\
0 & P & \xi & 0 \\
0 & \xi^* & Q & 0\\
 \varsigma^* & 0 & 0 & \chi
\end{array} \right),
\end{equation}\\
where $T_2$ is partial trace over second spin(note that the matrix is symmetric).
$\{\delta,\varsigma,P,Q,\xi,\chi\}$ are functions of $T$, $B$, $D$, $\gamma$, $J$ and $\zeta$, and also mixture of components of the total density matrix.

\section{Exact Numerical Solution}\label{numerical}
In order to provide a detailed analytical and numerical simulation, here, we use the concurrence as a measure of entanglement for the bipartite (sub)system, also CC is investigated and simulated numerically as well as the concurrence. 
\subsection{Correlation functions}
With regard to the geometric of correlation functions, we will calculate the concurrence and CC for the particular case bipartite spins (1/2,1/2) whose the density operator is presented in the form (\ref{density matrices}). Arrays of this matrix can be characterized as the following equations
\begin{equation}\label{CF}
\begin{array}{lcl}
\delta =\frac{1}{4}(1+\mathcal{G}^{i}_z+\mathcal{G}^{j}_z+\mathcal{G}^{ij}_{zz}),\quad
P =\frac{1}{4}(1+\mathcal{G}^{i}_z-\mathcal{G}^{j}_z-\mathcal{G}^{ij}_{zz}),\\
Q =\frac{1}{4}(1-\mathcal{G}^{i}_z+\mathcal{G}^{j}_z-\mathcal{G}^{ij}_{zz}),\quad
\chi =\frac{1}{4}(1-\mathcal{G}^{i}_z-\mathcal{G}^{j}_z+\mathcal{G}^{ij}_{zz}),\\
\varsigma =\frac{1}{4}(\mathcal{G}^{ij}_{xx}-\mathcal{G}^{ij}_{yy}),\quad
\xi =\frac{1}{4}(\mathcal{G}^{ij}_{xx}+\mathcal{G}^{ij}_{yy}),
\end{array}
\end{equation}
where $\mathcal{G}^{k}_z=\langle \sigma_{z}^k\rangle$ with $k=\{i,j\}$, is the magnetization density at site $k$ and $\mathcal{G}^{ij}_{\mu\nu}=\langle \sigma_{\mu}^i\sigma_{\nu}^j\rangle$ with $\mu,\nu=\{x,y,z\}$ denote spin-spin correlation functions at sites $i$ and $j$. Note that the expectation value can be defined as $Tr[\rho_{13}\mathcal{G}]$. If we introduce the elements
\begin{equation}\label{CFelement}
\begin{array}{lcl}
\mathcal{E}_1 = \xi +\xi^* +\varsigma +\varsigma^*,\quad
\mathcal{E}_2 = \xi +\xi^* -\varsigma +\varsigma^*,\\
\mathcal{E}_3 = \delta +\chi -P -Q,\quad
\mathcal{E}_4= \delta -\chi -P +Q,\quad
\mathcal{E}_5= \delta -\chi +P -Q,\quad
\end{array}
\end{equation}
in accordance with the reconstructed density matrix (\ref{density matrices}) as
\begin{equation}\label{CFelement}
\begin{array}{lcl}
\rho^{T2}_{13}=\frac{1}{4Z}\big(I\otimes I+\sum\limits_{i=1}^3\mathcal{E}_i(\sigma^i\otimes \sigma^i)+\mathcal{E}_4(I\otimes \sigma^3)+\mathcal{E}_5(\sigma^3\otimes I)\big),
\end{array}
\end{equation}
then, we obtain a simple equation for CC of the bipartite system represented in Eq. (\ref{CCEq}) as
\begin{equation}\label{CFelement}
\begin{array}{lcl}
CC(\rho_{\mathscr{A}\mathscr{B}})=\frac{(1-\mathcal{E})}{2}\log_2(1-\mathcal{E})+\frac{(1+\mathcal{E})}{2}\log_2(1+\mathcal{E}),
\end{array}
\end{equation}
in which $\mathcal{E}=\max\{|\mathcal{E}_1|,|\mathcal{E}_2|,|\mathcal{E}_3|\}$.
\subsubsection{Concurrence}
\begin{figure}
\begin{center}
\includegraphics[width=6cm,height=4cm]{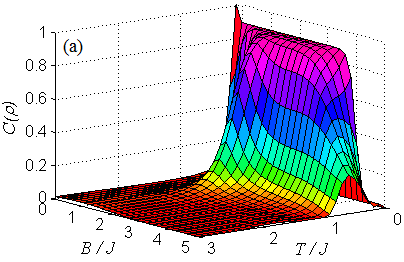}
\includegraphics[width=6cm,height=4cm]{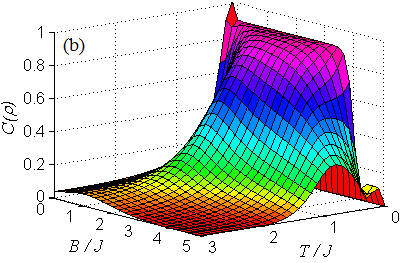}
\caption{Concurrence of the spins (1/2,1/2), with respect to the temperature and the magnetic field at fixed values of $\zeta=J$ and $D=5J$, for: (a) $\gamma=0.2J$; (b) $\gamma=0.8J$. }
\label{fig:concurrence1}
\end{center}
\end{figure}

 The reduced density matrix presented in Eq. (\ref{density matrices}) has whatever is needed about the bipartite spins (1/2,1/2), hence the concurrence can be readily obtained as \cite{Canosa}
 \begin{equation}\label{Newconcurrence}
\begin{array}{lcl}
\mathcal{C}(\rho_{\mathscr{A}\mathscr{B}})=2\max\{\max\big(0,|\frac{\varsigma}{Z}|-P\big),|\frac{\xi}{Z}|-(\frac{\delta \chi}{Z^2})^{\frac{1}{2}}\}.
\end{array}
\end{equation}
The concurrence (\ref{Newconcurrence}) as function of the temperature $T$ and the magnetic field $B$ at fixed values of the anisotropy, the single-ion anisotropy and DM interaction parameters is shown in Fig. \ref{fig:concurrence1}. As illustrated in this figure, the concurrence at low temperature and weak magnetic field is maximum ($\mathcal{C}(\rho)=1$),  on the other hand, this quantity is minimum ($\mathcal{C}(\rho)=0$) at high temperature and strong magnetic field. This essential property of the concurrence has been studied for various spin models in the previous works \cite{Wootters,Vedral,Cao3} and here it is true for our favorite bipartite (sub)system and is compatible with the previous works.

\begin{figure}
\begin{center}
\includegraphics[width=6cm,height=5cm]{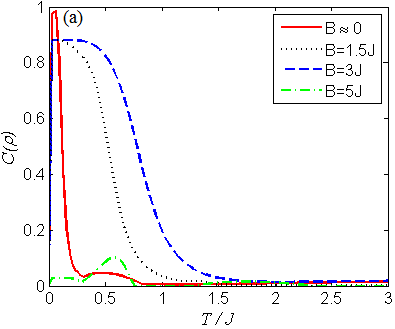}
\includegraphics[width=6cm,height=5cm]{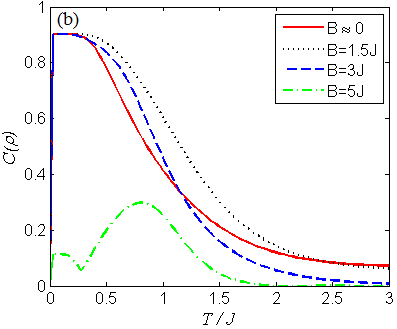}
\caption{Concurrence of the spins (1/2,1/2) as function of the temperature at fixed values of $\zeta=J$ and $D=5J$ in the various magnetic fields for: (a) $\gamma=0.2J$; (b) $\gamma=0.8J$.}
\label{fig:2DconcurrenceT}
\end{center}
\end{figure}

 In the some of used references, authors gained a critical temperature at which the concurrence vanishes, but here we see that with changes of the magnetic field, the concurrence vanishes at different critical temperatures. Indeed, by inspecting Fig. \ref{fig:2DconcurrenceT}, one can observe that for the various magnetic fields, the concurrence diagrams vanish in the various critical temperatures and that is because, the concurrence of the bipartite (sub)system strongly depends on the some extra parameters except the temperature and the magnetic field, such as the single-ion anisotropy and DM interaction parameters. 
Also, it is explicitly seen by increasing the anisotropy $\gamma$, the concurrence vanishes at higher critical temperatures for the fixed values of the magnetic field(blue dash line and green dot-dashed line). Meanwhile, at weak magnetic field, the concurrence not entirely be zero in the higher temperatures. In the stronger anisotropy $\gamma$, this phenomenon becomes more clear(Figs. \ref{fig:concurrence1}(b) and \ref{fig:2DconcurrenceT}(b)). As a result, because of verifying the entanglement for such quantum system at very low temperatures (near $T=0$) is practically difficult, it can be easier by increasing of the anisotropy. Indeed, one can use the anisotropy as an entanglement controller in higher temperatures for such model.
\begin{figure}
\begin{center}
\includegraphics[width=5.7cm,height=4.2cm]{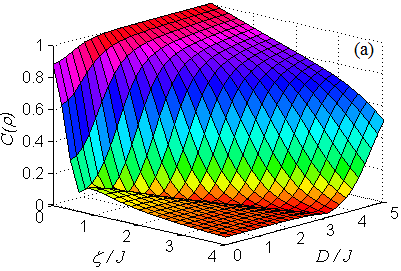}\quad
\includegraphics[width=5.7cm,height=4.2cm]{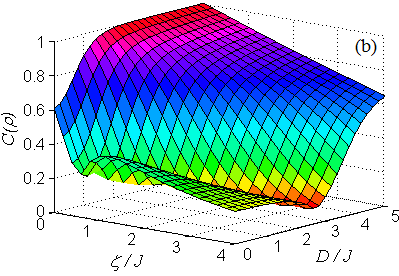}
\caption{Concurrence of the spins (1/2,1/2), with respect to the single-ion anisotropy and DM interaction parameters at low temperature $T=0.15J$ and fixed $B=J$ for: (a) $\gamma=0.2J$; (b) $\gamma=0.8J$.}
\label{fig:ConcurrenceZetaD}
\end{center}
\end{figure}

Figure \ref{fig:ConcurrenceZetaD} represents the concurrence of the spins (1/2,1/2) with respect to the single-ion anisotropy and the DM interaction parameters at low temperature($T=0.15J$) and fixed homogeneous magnetic field $B=J$. As illustrated in this figure, for the fixed values of the single-ion anisotropy, with decrease of the DM interaction from its high values, the concurrence decreases until reaches a minimum in which a sudden change occurs in a special critical DM interaction. Here, state of the bipartite (sub)system will change, indeed a phase transition occurs in this critical point. with increase of $\zeta$ this critical point tends to the stronger DM interaction. 

If we consider a line that connects these critical points, we see that the concurrence increases with increase of the DM interaction for the region upper than this line, namely, the concurrence is proportional to the DM interaction in this region. On the other hand, for the region lower than the connection line, the concurrence decreases with increase of the DM interaction.

\begin{figure}
\begin{center}
\includegraphics[width=6cm,height=5cm]{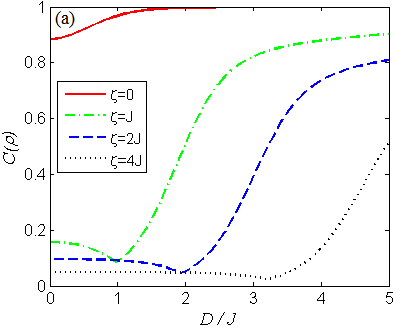}
\includegraphics[width=6cm,height=5cm]{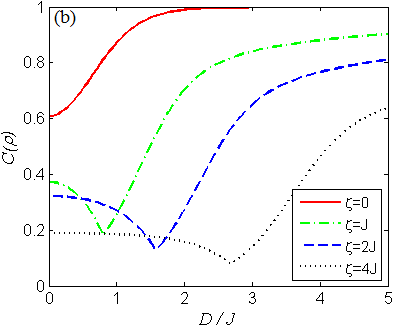}
\caption{ Concurrence of the spins(1/2,1/2) as function of the DM interaction parameter $D$ at low temperature $T=0.15J$ at fixed values of the single-ion anisotropy and $B=J$ for: (a) $\gamma=0.2J$; (b) $\gamma=0.8J$.}
\label{fig:2DConcurrenceD}
\end{center}
\end{figure}

\begin{figure}
\begin{center}
\includegraphics[width=6cm,height=5cm]{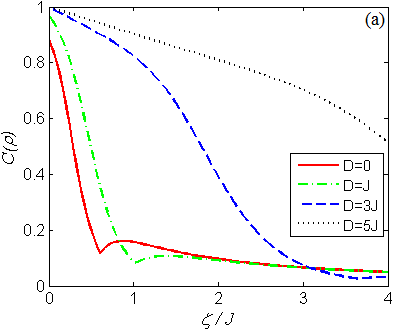}
\includegraphics[width=6cm,height=5cm]{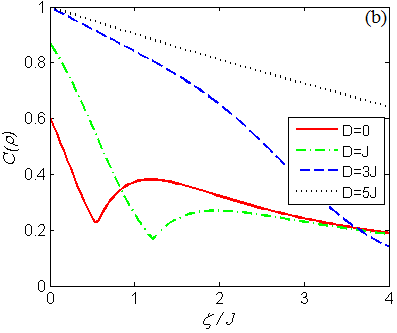}
\caption{Concurrence of the spins(1/2,1/2) as function of the single-ion anisotropy parameter $\zeta$ at low temperature $T=0.15J$ at fixed values of the DM interaction and $B=J$ for: (a) $\gamma=0.2J$; (b) $\gamma=0.8J$.}
\label{fig:2DConcurrenceS}
\end{center}
\end{figure}

 This function decreases with increase of the single-ion anisotropy $\zeta$ at fixed values of $D$ until reaches that minimum in which phase transition occurs as we mentioned before. Hence, the concurrence reaches its maximum value in the strong DM interaction and small single-ion anisotropy. Shape of the concurrence digram is changed by increasing the anisotropy $\gamma$. These properties are obviously presented in Figs. \ref{fig:2DConcurrenceD} and \ref{fig:2DConcurrenceS}.

As shown in Fig. \ref{fig:2DConcurrenceD}(a), maximum amount of the concurrence decreases with increase of the single-ion anisotropy $\zeta$. Moreover, in this figure it is obviously visible, that those critical points in which the concurrence reaches its minimum at low temperature tend to the stronger DM interaction, where $\zeta>0$ and $\gamma$ are considered fixed. For the strong single-ion anisotropy property $\zeta>4J$, we found that the minimum amount of the concurrence becomes zero, namely, in this condition state of the bipartite (sub)system is a separable state. With increase of the anisotropy parameter $\gamma$(from 0.2 to 0.8), the critical points shift to the weaker DM interaction(Fig. \ref{fig:2DConcurrenceD}(b)).

 In the present paper, we obliged ourselves to investigate the pairwise entanglement of spins (1/2,1/2) from the single-ion anisotropy point of view, which is merely considered for integer-spins in the total spin chain. Fig. \ref{fig:2DConcurrenceS} depicts the concurrence as function of the single-ion anisotropy $\zeta$ at fixed values of $D$ at low temperature. As illustrated in this figure, one can see that the pairwise entanglement sorely depends on the single-ion anisotropy $\zeta$. With increase of the single-ion anisotropy in the tripartite system, the concurrence of the bipartite (sub)system decreases for fixed values of $D$. 

 In the almost weak DM interaction(red  solid line and green dot-dashed line), the concurrence decreases by increasing the single-ion anisotropy from zero and reaches a minimum in which phase transition occurs. With further increase of the single-ion anisotropy, a sudden change happens in the concurrence behaviour at a special critical single-ion anisotropy $\zeta^c$ and for $\zeta>\zeta^c$ this quantity decreases smoothly. By increasing the anisotropy $\gamma$, this sudden change(the concurrence minimum point)  occurs at almost weaker the single-ion anisotropy for $D=0$(red solid  line in Fig. \ref{fig:2DConcurrenceS}), while for $D>0$ happens at higher single-ion anisotropy(green dot-dashed line). This behaviour is slightly complex but comprehensible and a reason that make our favorite system appealing to study. In the strong DM interaction, the concurrence decreases almost independent of the anisotropy $\gamma$ with increase of the single-ion anisotropy parameter $\zeta$(black dot line).
\begin{figure}
\begin{center}
\resizebox{1\textwidth}{!}{%
\includegraphics{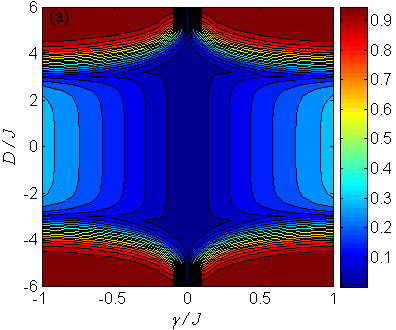}
\includegraphics{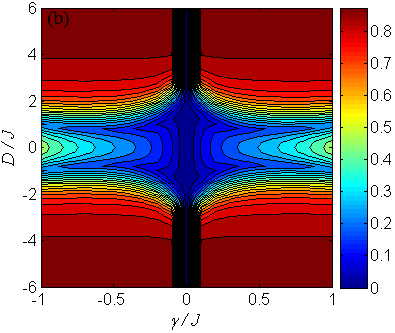}
}

\resizebox{1\textwidth}{!}{%
\includegraphics{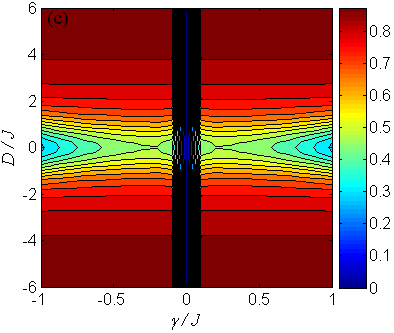}
\includegraphics{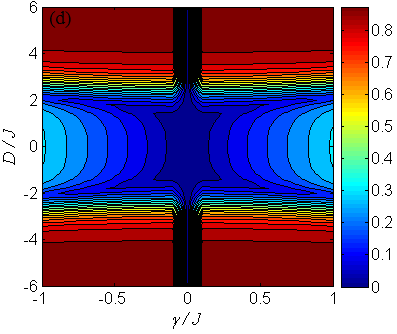}
}

\resizebox{1\textwidth}{!}{%
\includegraphics{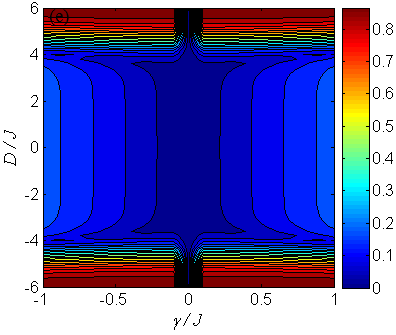}
\includegraphics{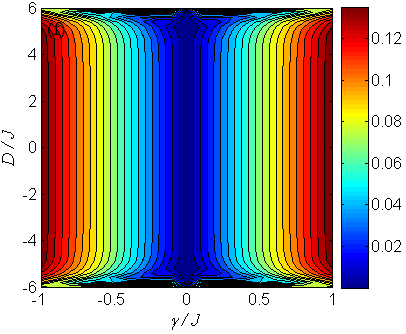}
}
\caption{Contour plots of the concurrence as function of the anisotropy $\gamma$ and the DM interaction $D$ at fixed values of $\zeta=J$ at low temperature $T=0.15J$ for: (a) $B=0$; (b) $B=J$; (c) $B=2J$; (d) $B=3J$; (e) $B=4J$; (f) $B=5J$, for spins (1/2,1/2). Colour bars represent changes of the concurrence from $C=0$(black regions) to $C=1$(red regions).}
\label{fig:RecontourConcurrenceDGamma}
\end{center}
\end{figure}

For better understanding the behaviour of the concurrence with respect to the anisotropy $\gamma$ and DM interaction, we knew interest that depict the anisotropy and the DM interaction dependences of the concurrence at low temperatures at fixed values of the magnetic field, in the form of contour plots as illustrated in Fig. \ref{fig:RecontourConcurrenceDGamma}. This figure shows that in this circumstances, the concurrence is symmetric versus the anisotropy and the DM interaction parameters for various magnetic fields. Also, it can be readily seen that in the strong DM interaction i.e., $D>|2J|$ for $0\leq B\leq2J$(Figs. \ref{fig:RecontourConcurrenceDGamma}(a)$-$\ref{fig:RecontourConcurrenceDGamma}(c)) and $D>|BJ|$ for $2J< B\leq4J$(Figs. \ref{fig:RecontourConcurrenceDGamma}(d)$-$\ref{fig:RecontourConcurrenceDGamma}(e)), the concurrence is maximum. In the presence of the strong magnetic field $B>4J$ the concurrence behaviour dramatically alters. We explain this exotic behaviour in the different intervals of the magnetic field in the following.

 If one follows Fig. \ref{fig:RecontourConcurrenceDGamma} step by step then realizes that the thermal concurrence as a measure of pairwise entanglement has an exotic behaviour versus increasing the magnetic field. This quantity for interval $0\leq B\leq 2J$ behaves different from interval $2J<B\leq 4J$, also its behaviour for $B>4J$ is generally different from former intervals. Namely, for the first interval, with increase of the magnetic field from zero to $2J$ the concurrence arises at the weaker DM interaction(the blue region gradually decreases) as far as the concurrence does not vanish even at $D=0$ and $\gamma\approx 0$. While, at the second interval, with increase of the magnetic field, the concurrence arises at stronger the DM interaction($D\approx |BJ|$), which width changes of the blue region can present this exotic behaviour. Consequently, for the strong magnetic field, the field influence on the concurrence behaviour  will overcome the DM interaction. Finally, with regard to Fig. \ref{fig:RecontourConcurrenceDGamma}(f) for $B>4J$ the maximum amount of the concurrence is entirely limited to the DM interaction interval $D<|BJ|$ and strong anisotropy $\gamma\approx |J|$.
 \begin{figure}
\begin{center}
\resizebox{1\textwidth}{!}{%
\includegraphics{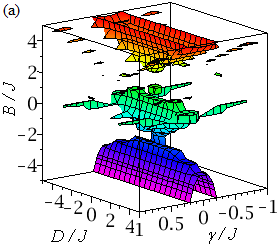}
\includegraphics{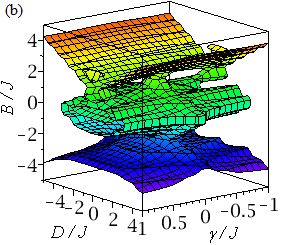}
\includegraphics{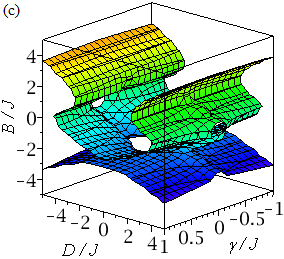}
}
\caption{ Surfaces of constant geometric concurrence of state $\rho(AB)$ defined in Eq. (\ref{density matrices}) at low temperature $T=0.15$ for: (a) $\zeta=J$ and $C(\rho)=0.03$, (b) $\zeta=J$ and $C(\rho)=0.3$, (c) $\zeta=J$ and $C(\rho)=0.8$. It is clear that the geometry of the concurrence is symmetric versus absolute values of the DM interaction and the anisotropy, while there is not any symmetry versus the magnetic field axes.}
\label{fig:Geometry}
\end{center}
\end{figure}

In Fig. \ref{fig:Geometry}, we plot level surfaces of geometric concurrence of the state $\rho(AB)$ defined in Eq. (\ref{density matrices}) at low temperature and fixed value of $\zeta=J$.  Using this figure, one can recognize that in what regions seeks existence of the entangled states for the bipartite (sub)system. Obviously, surfaces of the constant geometric concurrence are symmetric versus DM interaction and anisotropy $\gamma$, while we can not find any symmetry versus the magnetic field axes. Selecting of this special model may be reson of this symmetry breaking and one can choose a model for which symmetry always be established.
\subsubsection{Classical correlation}
 With regard to Sec. \ref{introduction} and references therein, we start to explain the configuration of the CC between the spins (1/2,1/2) as well as its concurrence, for realizing some extra physical behaviours of the (sub)system which are rare in the other  investigated spin models in the previous works. We here verify this quantity which can be exist in the both classical and quantum systems as function of the various parameters then, compare it with the concurrence. Finally, we get some interesting outcomes.
 \begin{figure}
\begin{center}
\includegraphics[width=6cm,height=5.5cm]{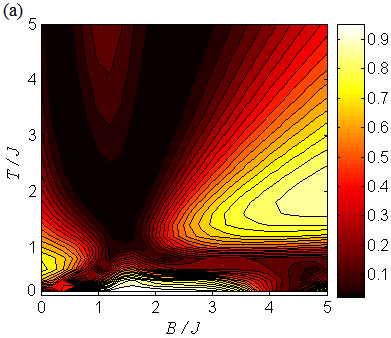}
\includegraphics[width=6cm,height=5.45cm]{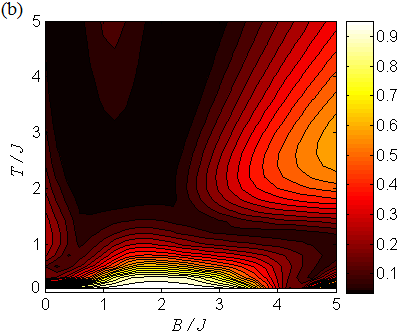}
\caption{ CC of the spins (1/2,1/2) as function of the temperature and the magnetic field at fixed values of $D=5J$ and $\zeta=J$ for: (a) $\gamma=0.2J$; (b) $\gamma=0.8J$.}
\label{fig:CCcontourBT}
\end{center}
\end{figure} 

 CC as function of the temperature and the magnetic field at fixed values of the single-ion anisotropy and the DM interaction parameters is shown in Fig. \ref{fig:CCcontourBT}. Let us divide this figure to four segments: (i) at low temperatures($T\ll J$) and interval $0.5J\lesssim B\lesssim 3.5J$, CC is maximum and in the outside of this interval this quantity vanishes. Also here, by increasing the temperature from zero this quantity gradually vanishes; (ii) at interval $0\leq B\lesssim 1$, by increasing the temperature from zero CC arises, and with further increase of the temperature until $T\approx 3J$, this quantity reaches a maximum value then vanishes; (iii) at  $B\gtrsim 2J$, by increasing the temperature from $T\approx 0.5J$, CC arises as far as at high temperatures this quantity reaches another maximum then vanishes again; (iv) for $T\gtrsim 3J$ at interval $0.5J\lesssim B\lesssim 1.5J$, with increase of the temperature, this quantity arises again and reaches a maximum smaller that other. It is clear that, with increase of the anisotropy the maximum value of CC decreases in the latest three region (ii), (iii) and (iv) but it increases in the first region (i).
 
If we look smartly at Fig. \ref{fig:concurrence1} from the top perspective, we realize that for low temperatures by increasing the anisotropy, the CC behaviour will almost becomes similar to the quantum entanglement for the spins (1/2,1/2). The subject matter is the behaviour of both functions at low temperature which almost become the same with increasing the anisotropy. Indeed, the anisotropy has the unification ability on the concurrence and CC functions at low temperature.

\begin{figure}
\begin{center}
\includegraphics[width=6cm,height=5cm]{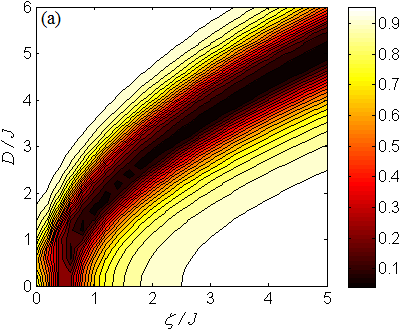}
\includegraphics[width=6cm,height=5cm]{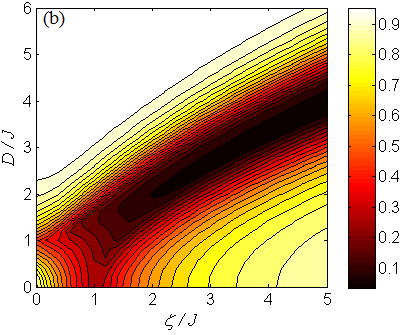}
\caption{ CC contour plots for the spins (1/2,1/2) as function of the single-ion anisotropy and the DM interaction at low temperature $T=0.15J$ and fixed $B=J$ for: (a) $\gamma=0.2J$; (b) $\gamma=0.8J$.}
\label{fig:CCcontourDZ}
\end{center}
\end{figure}

\begin{figure}
\begin{center}
\includegraphics[width=12cm,height=5.5cm]{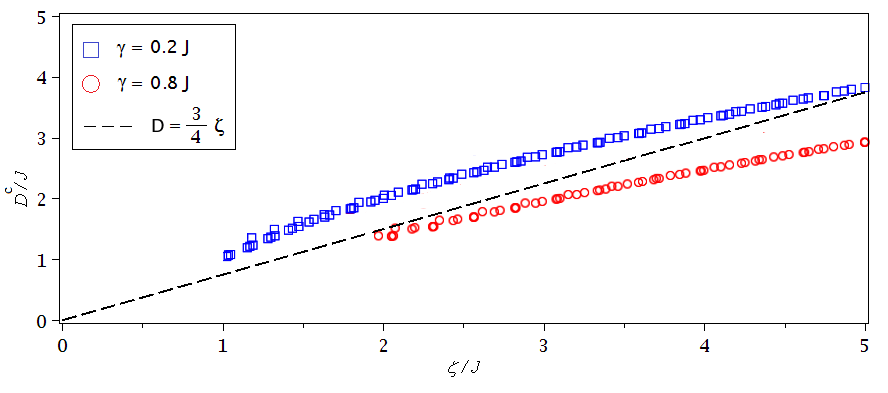}
\caption{ The critical DM interaction $D^c$ versus single-ion anisotropy $\zeta$ at low temperatures $T=0.15J$ and fixed $B=J$ for: (a) $\gamma=0.2J$(boxes); (b) $\gamma=0.8J$(circles). Note that the critical DM interaction points for case $\gamma=0.2J$ is above of the dashed line which represents linear equation $D=0.75\zeta$, while they are below of it for $\gamma=0.8J$.}
\label{fig:Boxed}
\end{center}
\end{figure}

 CC as function of the single-ion anisotropy and the DM interaction at low temperatures and fixed value of the magnetic field is shown in Fig. \ref{fig:CCcontourDZ}. As shown in this figure, in the strong DM interaction($D\gtrsim 5J$), CC is maximum independent of the single-ion anisotropy changes. But for the weaker DM interaction, with increase of the single-ion anisotropy, this quantity vanishes at a special region, then arises again and reaches another maximum. Such region that presents a critical DM interaction domain versus the single-ion anisotropy parameter is depicted in Fig. \ref{fig:Boxed}.
This figure shows that by increasing the single-ion anisotropy at low temperatures and fixed magnetic field $B=J$, CC vanishes at higher critical DM interaction for both $\gamma=0.2J$ and $\gamma=0.8J$ where, for the case of $\gamma=0.2J$ these critical points are above line $D=0.75\zeta$(dashed line), but for the case of $\gamma=0.8J$, they are below this line. This is means that, for Fig. \ref{fig:CCcontourDZ}(a) we have $0.75\zeta<D^c<\zeta$ while, for Fig. \ref{fig:CCcontourDZ}(b) inequality $0.5\zeta<D^c<0.75\zeta$ is established(also here, there are some points at which CC is maximum at weak DM interaction and small the single-ion anisotropy). By comparing Fig. \ref{fig:CCcontourDZ}(a) with Fig. \ref{fig:2DConcurrenceD}(a) one can gain some likenesses and differences in behaviour of the concurrence and CC. For example, at fixed values of $\zeta>0$ with increase of the DM interaction from zero, CC decreases from a maximum and reaches a minimum value at a critical point. With further increase of the DM interaction, this function increases and reaches another maximum, just like the concurrence. By increasing $\zeta$ the critical point tend to the stronger the DM interaction. Unlike the concurrence, both CC maxima before and after the critical point are equivalent for $\zeta\geq 2J$. Another achievements can be obtained by comparing all represented figures corresponding to the concurrence and CC.

\subsection{Transmission rate}\label{rate}
Quantum transmission rate for the introduced protocol in Fig. \ref{fig:spinchain} can be obtained with regard to the quantum channel capacity of the memoryless channel map $\mathcal{M}[\rho]$. This capacity that was defined as the maximum amount of the quantum information reliably transmitted per use of the channel $\mathcal{M}$ (see Ref. \cite{Bennett}), was used for a spin chain channel in Ref. \cite{Rossini} and is given by
\begin{equation}\label{QC}
\mathcal{Q}(\eta)=\max\limits_{p\in [0,1]}\{H_2(\eta p)-H_2((1-\eta)p)\},
\end{equation}
where, $H_2(X)=-X\log_2(X)-(1-X)\log_2(1-X)$ is the dyadic Shannon entropy.

The rate of this channel for $\varepsilon$ uses in the time interval $T=\varepsilon \tau$ can be defined as
\begin{equation}\label{rate}
\mathcal{R}\equiv \lim\limits_{\varepsilon \rightarrow \infty}\frac{\varepsilon \mathcal{Q}(\eta)}{\varepsilon \tau}=\frac{\mathcal{Q}(\eta)}{\tau}.
\end{equation}

Assume a simple spin-1/2 Heisenberg model with general Hamiltonian $H_G$. The operator of the total $z$-component of the spin, given by $\sigma_{total}^z=\sum_{i}\sigma_{i}^z$ is conserved, namely $\big[\sigma_{total}^z,H_G\big]=0$. Hence, the Hilbert space $\mathcal{H}_G$ decomposes into invariant subspaces, each of which is a distinct orthogonal eigenvector of the operator $\sigma_{total}^z$. Here, the transfer amplitude $\eta$ in Eq. (\ref{QC}) is obtained using confined Hamiltonian $\mathcal{H}_G$.
So, perfect and efficient quantum state transfer is happened for the such quantum chain with $\mathcal{R}=1$.
But, for our suggested model with Hamiltonian (\ref{hamiltonian}) as a communication channel, we obtain the transfer amplitude $\eta$ using general Hamiltonian $H_G$ of the system, and prove that the general Hamiltonian of the such system with few body is a capable operator to investigate the transmission protocol. Hence, we focus on the maximum and minimum amount of the rate $\mathcal{R}$.

 For our model, the transfer amplitude $\eta$ is a sinusoidal function of $\tau$ just the same one in Ref. \cite{Rossini}, but with changeable period $\pi/2(1-iD)$, where $i=\sqrt{-1}$, namely
\begin{equation}\label{eta}
\eta=|\Upsilon_{13}(\tau)|^2=|\sin(2(1-iD)\tau |^2.
\end{equation}

\begin{figure}
\begin{center}
\resizebox{1\textwidth}{!}{%
\includegraphics{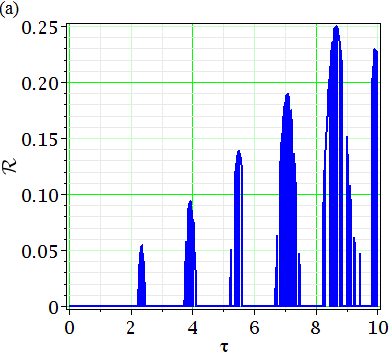}
\includegraphics{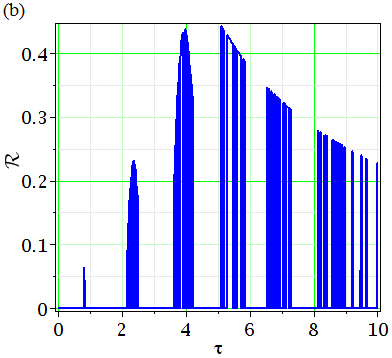}
}

\resizebox{1\textwidth}{!}{%
\includegraphics{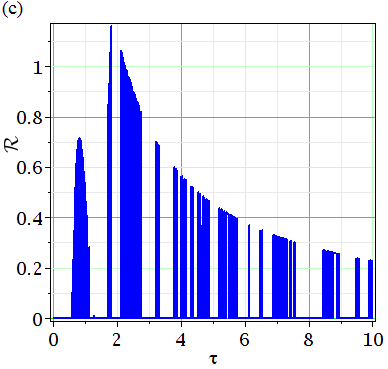}
\includegraphics{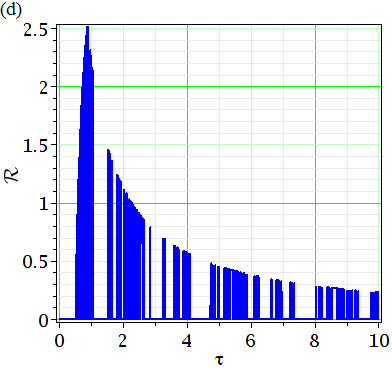}
}
\caption{Quantum transmission rate versus time $\tau$ for: (a) $D=0.05$; (b) $D=0.1$; (c) $D=0.3$; (d) $D=0.5$.}
\label{fig:Transferrate}
\end{center}
\end{figure}

By setting Eq. (\ref{eta}) in Eq. (\ref{QC}), we can get the quantum transmission rate (\ref{rate}). Figure \ref{fig:Transferrate} depicts  this rate versus time evolution $\tau$ numerically for various fixed values of the DM interaction. With regard to this figure, information transferring rate roughly depends on the DM interaction, i.e., for $D=0.05$(Fig. \ref{fig:Transferrate}(a)) with the pass of time, maximum transmission rate increases between time interval $0<\tau<10$ and reaches a biggest peak which represents maximum quantum transmission rate. By increasing  the DM interaction $D$ from 0.05 to 0.5 the biggest peak of the rate occurs with the pass of less time and its intensity increases almost ten times(from $\mathcal{R}=0.25$ to $\mathcal{R}=2.5$).

With regard to this protocol, we can realize that an unknown qubit can be reliably transferred through a mixed-three-spin (1/2,1,1/2) channel which spins (1/2,1/2) have both XY and DM interactions together, without considering spin-integer. Namely, the transmission rate is independent of the Ising coupling between spins (1,1/2) and single-ion anisotropy parameter related to the spin-integer. But as noted, it is roughly dependent on the DM interaction. If one looks carefully at the Fig. \ref{fig:Transferrate} and follows limited time interval $0<\tau<1$ then, can realize that with increase of the DM interaction, the transmission rate of the channel arises gradually and then reaches its  maximum value in this interval.

\section{Summary and conclusions}\label{conclusions}
 We have introduced a mixed-$N$-spin Ising-XY model, then focused on a mixed-three-spin (1/2,1,1/2) cell of it to investigate CC and quantum entanglement between spins (1/2,1/2) in the vicinity of a homogeneous magnetic field. Here, both XY and DM interactions between spins (1/2,1/2) also Ising interaction between spins (1,1/2) and a single-ion anisotropy for spin-integer have been considered. In this paper, we restricted ourselves to a finite chain with mixed-three-spin (1/2,1,1/2) of a larger mixed-$N$-spin chain, and because it is generally difficult to study a spin chain with large length, small-size systems can be  good options for obtaining some information about large-size systems. Fortunately, small-size cells of a large-size spin chain can also depict well properties of the large-size chain such as the temperature, the magnetic(here the DM interaction and the single-ion anisotropy for the suggested triangular cell) and the quantum correlation properties.
 
 In conclusion, we found that in the fixed values of $\zeta=J$ and $D=5J$, the concurrence is maximum at low temperature and weak magnetic field, but with increase of the temperature this quantity decreases until vanishes at a critical temperature. This critical temperature is changed with the magnetic field changes, namely, for various magnetic fields we have different critical temperatures. With increase of the anisotropy, these critical points shift to the higher temperatures. 
 
 Also, at low temperature and fixed value of $B=J$, the concurrence is maximum at strong DM interaction and zero single-ion anisotropy, and by increasing the single-ion anisotropy also decreasing the DM interaction this quantity decreases until reaches  a minimum at which phase transition occurs. This is means that the concurrence as a measure of entanglement associated to the half-spins, is roughly dependent on the single-ion anisotropy related to the spin-1. Some critical pints of the DM interaction and the single-ion anisotropy have been numerically presented in which the concurrence suddenly changes.
 
Moreover, at low temperatures, we investigated the concurrence as function of the DM interaction and anisotropy at various fixed values of the magnetic field, and realized that this quantity is symmetric in the DM interaction and the anisotropy framework and will extremely change with increase of the magnetic field. As a result, the concurrence has a different behaviour for field intervals $0\leq B\leq2J$, $2J< B\leq4J$ and $B>4J$.

Surfaces of constant geometric concurrence of the (sub)system (1/2,1/2) with state $\rho(AB)$ at low temperature and fixed $\zeta=J$ have been presented in this work. We concluded that these surfaces depict the position of regions contained the entangled states properly. By using these surfaces we can easier detect circumstances in which the (sub)system state is entangled.

In that follows, we verified CC for the bipartite spins (1/2,1/2), and understood that at fixed values of $\zeta=J$ and $D=5J$ this term has a exotic behaviour in the temperature and the magnetic field framework. As an interesting outcome, we showed that at low temperatures, by increasing the anisotropy, the behaviour of CC will almost be same as the concurrence. Also, this term is investigated at low temperature and fixed magnetic field $B=J$ with respect to the DM interaction and the single-ion anisotropy. Hence, we gained a set of the critical DM interaction in which CC is vanished, and by increasing the anisotropy $\gamma$ range of this set of the critical points is changed in the DM interaction and the single-ion anisotropy framework(Fig. \ref{fig:Boxed}).

Finally, we theoretically considered the tripartite mixed-spins (1/2,1,1/2) as a communication channel with capacity, then analyzed quantum information transmission rate of it. To perform the transmission protocol, we assumed that an unknown qubit can be reliably transferred through this channel. Here, among the introduced parameters for the favorite system, the transmission rate is just dependent on the DM interaction and that is because of the special Ising-XY considered model for the spin chain which is not mentioned in the previous works. As another interesting result, we showed that by increasing the DM interaction between spins (1/2,1/2), maximum transmission rate occurs at the less time interval and this channel without considering the spin-1, can reliably transfer quantum information.

One can compute another properties of the channel such as amplitude damping channel, entanglement-assisted classical capacity of
the channel etc. for our suggested model and obtains some interesting outcomes.
{

\end{document}